\documentclass[12pt]{article}
\begin{document}
\begin{center}
{\bf DYNAMICAL SYMMETRY BREAKING AND EFFECTIVE LAGRANGIANS IN
$U(n)$ FOUR-FERMION MODELS ($n=2,3$)}\\
\vspace{5mm}
 S.I. Kruglov \\
\vspace{5mm}
\textit{International Educational Centre, 2727 Steeles Ave. W, \# 104, \\
Toronto, Ontario, Canada M3J 3G9,\\ E-mail:
skrouglo@utm.utoronto.ca}
\end{center}

\begin{abstract}
The formation of scalar condensates and dynamical symmetry
breaking in the $U(n)$ four-fermion models (for $n=2,3$) with two
coupling constants has been studied by the functional integration
method. The bosonization procedures of the models under
consideration are performed in one loop approximation. The
propagators of fermions, collective Bose-fields (bound states of
fermions), as well as  the mass formulas for the fermions and
bosons, are found. It is shown that a self-consistent
consideration of four-fermion models, in the framework of
dimension regularization, provides explicit mass relations for
fermions (see equation (22)) for the case $n\geq3$. The effective
Lagrangian of interacting scalar bosons is also derived for the
case $n=2$.

\end{abstract}

\newpage

\section{Introduction}
Current research aims to derive the effective quark-meson
Lagrangians from the fundamental quantum chromodynamics (QCD)
Lagrangian, because QCD is the true theory of strong interactions
of quarks and gluons. When the perturbation theory is valid due to
the smallness of coupling constant $\alpha_s$, the predictions of
QCD at high energy levels are confirmed by experimental data.
However, reformulation of QCD at low energy levels (the coupling
constant $\alpha_s$ is not small) in terms of hadronic fields as
bounded quarks, meets serious difficulties as the nonperturbative
effects play a very important role. One of the mathematical
difficulties is the impossibility of integrating the generating
functional for Green's functions over gluonic fields, as the
corresponding integral is not a Gaussian path integral. QCD, at
low energy, can be described by local effective chiral Lagrangians
(ECL) \cite{Weinberg}, \cite{Gasser}, \cite{Buchvostov}. However,
ECL contain many free parameters.

Under low energy quarks have approximately a contact four-fermion
interaction \cite{Klevansky}, \cite{Volkov}. Models with
four-fermion interactions are similar to the model of
superconductivity. Such models take into account both quarks and
mesons \cite{Eguchi}. The problem of renormalization and dynamic
mass formation in a four-fermion model with scalar-scalar,
pseudoscalar-pseudoscalar and vector-vector interactions was
investigated in \cite{KruglovActa}. A $CP$-odd, chirally
noninvariant, four-fermion model with three coupling constants was
studied in \cite{Kruglov89}.

In the theory of instanton vacuum \cite{Shuryak}, \cite{Dyakonov},
the chiral condensate appears to lead to breaking of the symmetry
and an effective four-fermion interaction thus emerges
\cite{Hooft} (see also \cite{Kruglov90}). It was noted by authors
\cite{Nambu} that a phenomenon of chiral symmetry breaking (CSB)
occurs in four-fermion models due to the self-interaction of
fermion fields. The vacuum is reconstructed in models considered,
and $\gamma_5 $ - symmetry is broken. As a result, the condensate
is formed and the condition for a vacuum expectation $\langle
\bar{\psi}\psi\rangle \neq 0$ is valid, and fermions acquire
masses $m=-g_0 \langle \bar{\psi}\psi \rangle $. It should be
noted that low masses of pions can be explained by CSB
\cite{Nambu60}.

The nonperturbative effects of CSB \cite{KruglovTMF} and a
confinement of quarks play a very important role in strong
interactions. Thus, four-fermion models describe CSB perfectly,
but not the confinement of quarks. I mention the QCD string
approach \cite{Simonov} (see also \cite{Kruglov}), which takes
into consideration the nonperturbative effect of quark
confinement. To include the confinement of quarks, one may
introduce the additional nonlocal quark interactions.

Here we consider the four-fermion models with the internal groups
$U(n)$ ($n=2,3$) which can be identified with the flavor group of
quarks but without including the colour group of symmetry. It
should be noted that the symmetry group of strong interactions is
the chiral $SU(n)\bigotimes SU(n)$ group. We investigate formal
four-fermion models with the $U(n)$ group (which are chiral
non-invariant models) and two coupling constants. The
consideration of such models allows us to concentrate on the
phenomena of dynamical symmetry breaking (DSB) which leads to the
mass formation. At the same time, it is not difficult to expand
the models under consideration thus introducing the additional
colour symmetry (terms recovering chiral symmetry etc.) that do
not influence the DBS, vacuum condensates or the mass formation.

 The paper is organized as follows: In Sec. 2, the
dynamical mass formation of fermions and the $U(n)$ ($n=2,3$)
symmetry breaking in scalar-scalar four-fermion model with two
coupling constants are considered. The bosonization procedure of
the model under consideration is performed. Masses and propagators
of collective bosonic fields (bound states of fermions) are
derived in Sec. 3. In Sec. 4, the effective Lagrangian of
interacting bosonic fields is evaluated for the case of the $U(2)$
group. Sec. 5 discusses results.

\section{Dynamical mass formation and symmetry breaking}

Consider a model with the internal symmetry group $U(n)$ and two
coupling constants \cite{Kruglov84}, \cite{Kruglov85}:
\begin{equation}
{\cal L }(x)=- \overline{\psi}(x)(\gamma_\mu \partial_\mu +m)\psi
(x) +\frac{F}{2}\left[ \overline{\psi}(x)\psi (x)\right]^2
+\frac{G}{2}\left[ \overline{\psi}(x)T^a\psi (x)\right]^2  ,
\label{1}
\end{equation}
where $T^a$ ($a=1,...,n^2-1$) are the generators of the group
$SU(n)$, $\partial_\mu =(\partial/\partial x_i
,-i\partial/\partial x_0)$ ($x_0$ is the time), $m$ is the bare
mass of fermions, $\gamma_\mu$ are the Dirac matrices,
$\psi=(\psi_1 , \psi_2 , ..., \psi_n )$ is the multiplet of
fermions. For the $SU(2)$ group we use doublet of fermions $\psi$
and the generators $T^a\equiv\tau ^a$ ($\tau^a$ are the Pauli
matrices, $a=1,2,3$), and for the $SU(3)$ group -- triplet of
fermions $\psi$ and the generators $T^a\equiv\lambda^a$
($\lambda^a$ are the Gell-Mann matrices $a=1,2,...,8$). We took
into consideration only scalar-scalar interactions which lead to
DSB. It is convenient to investigate DSB and the mass formation
with the help of a functional integration method \cite{Pervushin}.

The generating functional for Green's functions
\begin{equation}
Z[ \overline{\eta},\eta]=N_0\int D \overline{\psi}D\psi
\exp\left\{i\int d^4 x\left[{\cal L}(x)+  \overline{\psi}(x)\eta
(x)+ \overline{\eta}(x)\psi (x)\right]\right\} , \label{2}
\end{equation}
where $ \overline{\eta}$, $\eta$ are external sources,
$D\psi=\prod_x d\psi(x)$, using the replacement
\[
N_0=N\int D\Phi_0 D\Phi_a \exp \biggl\{ -i \int d^4 x
\biggl[\frac{\mu^2}{2}\left(\Phi_a (x)-\frac{g}{\mu^2}
\overline{\psi}(x)T^a \psi (x)\right)^2
\]
\[
+\frac{M^2}{2}\left(\Phi_0 (x)-\frac{f}{M^2} \overline{\psi}
(x)\psi (x)\right)^2 \biggr]\biggr\}
\]
can be represented as
\[
Z[\overline{\eta},\eta]=N\int D \overline{\psi}D\psi D\Phi_0
D\Phi_a \exp\biggl\{i\int d^4
x\biggl[-\overline{\psi}(x)\biggl[\gamma_\mu
\partial_\mu +m -f\Phi_0 (x)
\]
\vspace{-7mm}
\begin{equation} \label{3}
\end{equation}
\vspace{-7mm}
\[
-g\Phi_a (x) T^a\biggr]\psi (x)-\frac{\mu^2}{2}\Phi_a^2 (x) -
\frac{M^2}{2}\Phi_0^2 (x) +\overline{\psi}(x)\eta (x)
+\overline{\eta}(x)\psi (x) \biggr]\biggr\} .
\]
 Constants $F=f^2/M^2$, $G=g^2/\mu^2$ are introduced here, where
 $f$, $g$ are dimensionless constants and the constants $M$,
 $\mu$ have the mass dimensionality. So, dimensional constants $M$, $\mu$ play the
 role of bare masses of bosonic fields $\Phi_0 (x)$, $\Phi_a (x)$ and dimensionless constants $f$, $g$
 are bare coupling constants. The integral in Eq.~(3) is Gaussian in Fermi fields, and
 after integrating over the $\overline{\psi}$, $\psi$, we obtain
\begin{equation}
Z[\overline{\eta},\eta]=N\int D\Phi_0 D\Phi_a \exp\biggl\{iS[\Phi]
+ i\int d^4 xd^4 y \overline{\eta}(x)S_f (x,y)\eta(y)\biggr\} ,
\label{4}
\end{equation}
\[
S[\Phi]=-\frac{1}{2}\int d^4 x\left[M^2 \Phi_0^2 (x)+\mu^2
\Phi_a^2 (x)\right]
\]
\vspace{-7mm}
\begin{equation} \label{5}
\end{equation}
\vspace{-7mm}
\[
-i \mbox{tr} \ln \left[1+\widehat{G}\left(f\Phi_0 (x)+g\Phi_a
(x)T^a\right) \right] ,
\]
where $S[\Phi]$ is the effective action for bosonic collective
fields $\Phi_0 (x)$, $\Phi_a (x)$, and Green's functions
$\widehat{G}$, $S_f (x,y)$ for free fermions and for fermions in
the external collective fields obey the equations
\begin{equation}
(\gamma_\mu \partial_\mu +m)\widehat{G}(x,y)=-\delta(x-y) ,
 \label{6}
\end{equation}
\begin{equation}
\left[\gamma_\mu \partial_\mu +m-f\Phi_0 (x)-g\Phi_a
(x)T^a\right]S_f (x,y)=\delta(x-y) .
 \label{7}
\end{equation}
In order to formulate the perturbation theory \cite{Pervushin}, we
should find a solution of Eq.~(7) for the vacuum averages of the
fields $\Phi_0$, $\Phi_a$ independent of coordinates (the case of
the mean field approximation). In the momentum space Eq.~(7) takes
the form
\begin{equation}
(i\widehat{p} -A)S_f (p)=1 ,
 \label{8}
\end{equation}
where $\widehat{p}=p_\mu \gamma_\mu$, $p_\mu=(\textbf{p},ip_0)$,
$A=-m+f\Phi_0+g\Phi_a T^a$. According to Hamilton-Cayley theorem,
the matrix $A$ satisfies its characteristic equation:
\begin{equation}
A^2 -b_1 A+\det A=0 \hspace{0.5in}\mbox{for}\hspace{0.3in} U(2) ,
 \label{9}
\end{equation}
\begin{equation}
A^3-b_1A^2 +b_2 A-\det A=0\hspace{0.5in}\mbox{for}\hspace{0.3in}
U(3) ,
 \label{10}
\end{equation}
where
\[
b_1 =\mbox{tr}A ,\hspace{0.5in}b_2
=\frac{1}{2}\left[\left(\mbox{tr}A\right)^2-\mbox{tr}\left(A^2\right)\right]
.
\]
Let us search for a solution to Eq. (8) in the form
\begin{equation}
S_f (p)=a+b\widehat{p} +c_i A^i+d_i \widehat{p}A^i ,
 \label{11}
\end{equation}
where $i=1$ for the $U(2)$ group and $i=1,2$ for the $U(3)$ group,
$\widehat{p}A^i$ has the meaning of the direct product of the
matrices $\widehat{p}$ and $A^i$, $A^i$ is the $i$-th power of the
matrix $A$, and a summation over $i$ (for the $U(3)$ group) is
assumed. Substituting Eq.~(11) into Eq.~(8), with the help of
Eqs.~(9), (10) and the fact that the matrices $I$ (unit matrix),
$\widehat{p}$, $A^i$, $\widehat{p}A^i$ are linearly independent,
we obtain a system of equations for the unknown coefficients.
Solving this system, we obtain for the $U(2)$ group
\[
a=-\frac{b_1 \det A}{\Delta_1} ,\hspace{0.3in}
b=-\frac{i}{\Delta_1}\left(p^2 -\det A +b_1^2\right)
,\hspace{0.3in} d_1 =\frac{ib_1}{\Delta_1} ,
 \]
\begin{equation}
c_1 =-\frac{1}{\Delta_1}\left(p^2 -\det A\right) ,\hspace{0.3in}
\Delta_1 =\left(p^2 +m_1^2 \right)\left(p^2 +m_2^2 \right)
,\label{12}
\end{equation}
\[
m_1=m-f\Phi_0 -g \sqrt{\Phi_a^2} ,\hspace{0.3in}m_2=m-f\Phi_0 +g
\sqrt{\Phi_a^2} ,
\]
and for the $U(3)$ group
\[
a=\frac{i \det A}{\Delta_2}\left(p^2 -b_2\right) ,\hspace{0.3in}
b=-\frac{i}{\Delta_2}\left[\left(p^2 -b_2\right)^2+b_1\left(p^2
b_1-b_1\det A \right)\right] ,
\]
\[
 c_1 =-\frac{1}{\Delta_2}\left[p^4 -\left(b_2 +b_1^2\right)p^2 -b_1
 \det A\right] ,\hspace{0.3in}c_2 =\frac{1}{\Delta_2}\left(b_1^2 p^2 -
 \det A\right) ,
 \]
\begin{equation}
 d_1 =-\frac{i}{\Delta_2}\left(\det A -b_1 b_2 \right) ,
\hspace{0.3in}d_2 =\frac{i}{\Delta_2}\left(p^2 - b_2 \right) ,
\label{13}
\end{equation}
\[
 \Delta_2 =\det \left(p^2 +A^2\right)= p^2 \left(p^2 -b_2\right)^2+\left(
 \det A-p^2b_1 \right)^2
 \]
 \[
=\left(p^2 +m_1^2 \right)\left(p^2 +m_2^2 \right)\left(p^2 +m_3^2
\right) .
\]
The eigenvalues of the Hermitian matrix $(-A)$ determine the real
dynamical masses of the fermions (the spectrum mass). If the bare
masses of fermions $m$ are zero ($m=0$), they acquire the
different masses because of DSB. The expressions (11)--(13) define
the fermionic Green function in a covariant form, since all the
coefficients are expressed through the invariants of the $U(n)$
($n=2,3$) group. It is convenient to choose the gauge in which the
matrix $A$ is diagonal. In this case, we can put $\Phi_0\neq 0$,
$\Phi_3\neq 0$, $\Phi_8\neq 8$ (for $U(3)$), setting the rest of
$\Phi_a$ to zero. Green's function (11) then takes a diagonal form
\begin{equation}
S_f (p)=\left(
\begin{array}{cccc}
\frac{-i\widehat{p}+m_1}{p^2 +m_1^2} & . & 0 \\
. & . & . \\
0 & . & \frac{-i\widehat{p}+m_n}{p^2 +m_n^2} &
\end{array}
\right)  ,
 \label{14}
\end{equation}
where the masses of fermions are
\[
m_1=m-f\Phi_0 -g \Phi_3 ,\hspace{0.3in}m_2=m-f\Phi_0 +g \Phi_3
\hspace{0.3in}\mbox{for}\hspace{0.3in} U(2) ,
\]
\begin{equation}
m_1=m-f\Phi_0 -g \Phi_3-\frac{g}{\sqrt{3}}\Phi_8
,\hspace{0.3in}m_2=m-f\Phi_0 +g \Phi_3-\frac{g}{\sqrt{3}}\Phi_8 ,
\label{15}
\end{equation}
\[
 m_3=m-f\Phi_0 +\frac{2g}{\sqrt{3}}\Phi_8 \hspace{0.5in}\mbox{for}
 \hspace{0.3in} U(3) .
\]
It follows from this that if the fermion bare masses $m=0$, the
fermions still acquire nonvanishing dynamical masses. From
Eqs.~(15) we find the values of condensates $\Phi_0$, $\Phi_3$,
$\Phi_8$ (for the case of the $U(3)$ group) via fermion masses:
\[
2g\Phi_3=m_2-m_1 ,\hspace{0.3in}3\left(m-f\Phi_0
\right)=m_1+m_2+m_3 ,
\]
\vspace{-7mm}
\begin{equation} \label{16}
\end{equation}
\vspace{-7mm}
\[
2\sqrt{3}g\Phi_8 =2m_3-m_1-m_2 .
\]
It is seen from Eq.~(16) that the bare masses of fermions $m$ are
absorbed by the vacuum field $\Phi_0$.

\section{Masses and propagators of collective bosonic fields}

In order to obtain the vacuum condensates $\Phi_0$, $\Phi_3$,
$\Phi_8$ from Eq.~(5), we solve the equations for the fields
$\Phi_A (x)$ ($A=0,1,...,8$):
\[
\frac{\delta S[\Phi]}{\delta\Phi_0 (x)}=-M^2\Phi_0
(x)+if\mbox{tr}S_f (x,x)=0 ,
\]
\vspace{-7mm}
\begin{equation} \label{17}
\end{equation}
\vspace{-7mm}
\[
\frac{\delta S[\Phi]}{\delta\Phi_a (x)}=-\mu^2\Phi_a
(x)+ig\mbox{tr}\left[S_f (x,x)T^a\right]=0 .
\]
Substituting Eq.~(14) into Eq.~(17), we obtain a system of
equations (gap equations) for the vacuum averages (for the $U(3)$
group):
\[
M^2\Phi_0 =f\left(I_1m_1 +I_2m_2 +I_3m_3\right) ,
\]
\begin{equation}
\mu^2\Phi_3 =g\left(I_1m_1 -I_2m_2\right) ,  \label{18}
\end{equation}
\[
\sqrt{3}\mu^2\Phi_8 =g\left(I_1m_1 +I_2m_2-2I_3m_3\right) ,
\]
where
\begin{equation}
I_j =\frac{i}{4\pi^4}\int \frac{d^4 p}{p^2
+m_j^2}\hspace{0.5in}(j=1,2,3,\hspace{0.3in}d^4 p=id^3 pdp_0 ) .
\label{19}
\end{equation}
In considering the $U(2)$ group, it is necessary to take into
consideration the first two equations in (18) and to put $m_3=0$.
Due to the phase transitions massless fermions ($m=0$) become
massive. The massive states of fermions correspond to the minimum
of effective potential \cite{Jona}, \cite{Coleman}.

The integrals in Eq.~(19) are quadratically divergent and we can
use dimensional regularization \cite{Hooft72} or the
momentum-cutoff $\Lambda$ which specifies the region of nonlocal
interactions of fermions. Note that with cutoff regularization,
gap equations (18) have non-trivial, non-analytic solutions
\cite{Nambu} if $F\Lambda^2>2\pi^2$, $G\Lambda^2>2\pi^2$. The
integrals with cutoff ($\Lambda$) regularization play the role of
form-factors. The parameter $\Lambda$ defines the region of
non-locality of quark-antiquark forces. When
$\Lambda\rightarrow\infty$, four-fermion interactions become local
interactions.

It is known that dimensional regularization is most suited for
preserving the symmetry properties of the model. With the help of
the dimensional regularization quadratic and logarithmic divergent
integrals are given by \cite{Ramond}:
\[
\int \frac{d^{4-2\varepsilon}p}{p^2+m^2}=i\pi^{2-\varepsilon}
\Gamma(\varepsilon-1)\left(m^2\right)^{1-\varepsilon} ,
\]
\vspace{-7mm}
\begin{equation} \label{20}
\end{equation}
\vspace{-7mm}
\[
\int
\frac{d^{4-2\varepsilon}p}{\left(p^2+m^2\right)^2}=i\pi^{2-\varepsilon}
\Gamma(\varepsilon)\left(m^2\right)^{-\varepsilon} ,\hspace{0.3in}
\Gamma(\varepsilon)=(\varepsilon-1)\Gamma(\varepsilon-1),
\]
where $\Gamma(x)$ is the gamma-function and $\varepsilon$ is the
parameter of the dimensional regularization (we use the notation
$dp_4 =idp_0$). In this case, we come to the constraint (see
Eq.~(19))
\begin{equation}
I_i =\left(\frac{m_i}{m_j}\right)^2 I_j\hspace{0.5in}(i,j=1,2,3) .
\label{21}
\end{equation}
Using Eq.~(21), we find from Eq.~(18) the relation $\Phi_3
/(\sqrt{3}\Phi_8 )=(m_1^3 -m_2^3 )/(m_1^3+m_2^3-2m_3^3 )$.
Comparing this relation with the equality $\Phi_3 /(\sqrt{3}\Phi_8
)=(m_1 -m_2 )/(m_1+m_2-2m_3 )$, which follows from Eq.~(16), we
arrive at the mass formula (for the case of the $U(3)$ group
only):
\begin{equation}
\left(m_1-m_2\right)\left(m_1-m_3\right)\left(m_2-m_3\right)\left(m_1
+m_2 +m_3\right)=0 . \label{22}
\end{equation}
It follows from Eqs.~(22), (15) that there are solutions as
follows: 1) $m_1=m_2$ that is $\Phi_3 =0$; 2) $m_1=m_3$ that
requires $\sqrt{3}\Phi_8 =-\Phi_3$; 3) $m_2=m_3$ or
$\sqrt{3}\Phi_8 =\Phi_3$;  4) $m_1+m_2 +m_3=0$ that is equivalent
to $m=f\Phi_0$. So, self-consistent consideration of gap equations
(18) and mass formulas (15) requires one of four conditions. The
last possibility implies the negative mass of the third fermion
$m_3 =-m_1 -m_2$ if $m_1>0$, $m_2>0$. If we use the
momentum-cutoff $\Lambda$, Eqs.~(21), (22) are not valid.

In order to formulate the perturbation theory, we expend the
fields $\Phi_A (x)$, $A=(0,a)$, in Eq.~(4) in the neighborhood of
the vacuum averages, these later being the solutions of Eqs.~(18):
\[
\Phi_0 (x)=\Phi_0 +\Phi_0 '(x),\hspace{0.3in} \Phi_3 (x)=\Phi_3
+\Phi_3 '(x) ,\hspace{0.3in} \Phi_8 (x)=\Phi_8 +\Phi_8 '(x) .
\]
After expanding the logarithm in Eq.~5 in the power of the fields
$\Phi_A'$, the effective action (5) can be represented as
\[
S[\Phi']=-\frac{1}{2}\int d^4 xd^4 y\Phi_A
'(x)\Delta^{-1}_{AB}(x,y)\Phi_B '(y) +
\sum_{n=3}^{\infty}\frac{i}{n}\mbox{tr}\left[S_f\left(f\Phi_0 '+
g\Phi_a 'T^a\right)\right]^n ,
\]
\vspace{-7mm}
\begin{equation} \label{23}
\end{equation}
\vspace{-7mm}
\[
\Delta^{-1}_{AB}(p)=-ig_A g_B \mbox{tr}\left[\int\frac{d^4
k}{(2\pi^4)} S_f (k)T_A S_f (k-p)T_B\right]+\delta_{AB}M^2_A ,
\]
where $g_A=(f,g)$, $M_A=(M,\mu)$, $T_A=(1,T^a)$.

Generators of the $SU(2)$ group are the Pauli matrices $\tau ^a$
($a=1,2,3$), and for the $SU(3)$ group -- Gell-Mann matrices
$\lambda^a$ ($a=1,2,...,8$). Calculating the nonvanishing elements
of the inverse propagators of the auxiliary fields $\Phi'_A (x)$
in the momentum space, with the accuracy of ${\cal O}(g^2)$,
${\cal O}(f^2)$, ${\cal O}(fg)$, one finds for the $U(2)$ group:
\[
\Delta_{11}^{-1}(p)=\Delta_{22}^{-1}(p)=\mu^2+g^2\left(I_1
+I_2\right)+\left[p^2+\left(m_1+m_2\right)^2\right]Z_3^{-1} ,
\]
\[
\Delta_{33}^{-1}(p)=\mu^2+g^2\left(I_1
+I_2\right)+\left[p^2+2\left(m_1^2+m_2^2\right)\right]Z_3^{-1} ,
\]
\vspace{-7mm}
\begin{equation} \label{24}
\end{equation}
\vspace{-7mm}
\[
\Delta_{00}^{-1}(p)=M^2+f^2\left(I_1
+I_2\right)+\left[p^2+2\left(m_1^2+m_2^2\right)\right]\frac{f^2}{g^2}Z_3^{-1}
,
\]
\[
\Delta_{03}^{-1}(p)=fg\left(I_1
-I_2\right)+\frac{2f}{g}\left(m_1^2-m_2^2\right)Z_3^{-1} ,
\]
and for the $U(3)$ group
\[
\Delta_{11}^{-1}(p)=\Delta_{22}^{-1}(p)=\mu^2+g^2\left(I_1
+I_2\right)+\left[p^2+\left(m_1+m_2\right)^2\right]Z_3^{-1} ,
\]
\[
\Delta_{33}^{-1}(p)=\mu^2+g^2\left(I_1
+I_2\right)+\left[p^2+2\left(m_1^2+m_2^2\right)\right]Z_3^{-1} ,
\]
\[
\Delta_{44}^{-1}(p)=\Delta_{55}^{-1}(p)=\mu^2+g^2\left(I_1
+I_3\right)+\left[p^2+\left(m_1+m_3\right)^2\right]Z_3^{-1} ,
\]
\[
\Delta_{66}^{-1}(p)=\Delta_{77}^{-1}(p)=\mu^2+g^2\left(I_2
+I_3\right)+\left[p^2+\left(m_2+m_3\right)^2\right]Z_3^{-1} ,
\]
\begin{equation}
\Delta_{00}^{-1}(p)=M^2+f^2\left(I_1+I_2
+I_3\right)+\left[p^2+\frac{4}{3}\left(m_1^2+m_2^2+m_3^2\right)\right]\frac{3f^2}{2g^2}Z_3^{-1}
, \label{25}
\end{equation}
\[
\Delta_{88}^{-1}(p)=\mu^2+\frac{g^2}{3}\left(I_1+I_2
+4I_3\right)+\left[p^2+\frac{2}{3}\left(m_1^2+m_2^2\right)+\frac{8}{3}m_3^2\right]Z_3^{-1}
,
\]
\[
\Delta_{03}^{-1}(p)=fg\left(I_1
-I_2\right)+\frac{2f}{g}\left(m_1^2-m_2^2\right)Z_3^{-1} ,
\]
\[
\Delta_{08}^{-1}(p)=\frac{fg}{\sqrt{3}}\left(I_1
+I_2-2I_3\right)+\frac{2f}{\sqrt{3}g}\left(m_1^2+m_2^2-2m_3^2\right)Z_3^{-1}
,
\]
\[
\Delta_{38}^{-1}(p)=\frac{g^2}{\sqrt{3}}\left(I_1
-I_2\right)+\frac{2}{\sqrt{3}}\left(m_1^2-m_2^2\right)Z_3^{-1} ,
\]
where the constant of renormalization is given by
\begin{equation}
Z_3^{-1} =-\frac{ig^2}{4\pi^4}\int \frac{d^4
q}{\left(q^2+m^2_1\right)^2} . \label{26}
\end{equation}
Let us introduce the renormalized fields
\[
\Phi_a (x)=Z_3^{-1/2}\Phi_a '(x) ,\hspace{0.3in}\Phi_0
(x)=\sqrt{\frac{n}{2}}\frac{f}{g}Z_3^{-1/2}\Phi_0 ' (x)
\]
and coupling constants $g'^2=Z_3 g^2$, $f'^2=Z_3 f^2$. Using the
dimensional regularization, Eqs.~(20), and the relation
\cite{Ramond} $\lim_{\varepsilon \rightarrow 0}\varepsilon
\Gamma(\varepsilon-1)=-1$, we arrive at the constraint (see also
\cite{Scadron}):
\begin{equation}
Z_3^{-1} =\frac{g^2}{m_1^2}I_1 - \frac{g^2}{4\pi^2} . \label{27}
\end{equation}
Up to ${\cal O}(g^2)$, ${\cal O}(f^2)$, we find the renormalized
free (quadratic in the collective fields) effective action
\begin{equation}
S_{free}[\Phi] =-\frac{1}{2}\int d^4 x\left[\left(\partial_\mu
\Phi_A (x)\right)^2+m^2_{AB}\Phi_A (x)\Phi_B (x)\right] ,
\label{28}
\end{equation}
where $A=(0,a)$ and the elements of the mass matrices for the
$U(2)$ group are given by
\[
m_{00}^2=3\left(m_1^2+m_2^2\right)+\frac{2\left(m_1^3+m_2^3\right)}{2m-m_1-m_2}
,\hspace{0.3in}m_{11}^2=m_{22}^2=0 ,
\]
\vspace{-7mm}
\begin{equation}\label{29}
\end{equation}
\vspace{-7mm}
\[
m_{03}^2=3\left(m_1^2-m_2^2\right)
,\hspace{0.3in}m_{33}^2=\left(m_1-m_2\right)^2 .
\]
We took into consideration that according to the gap equations
(18) for the $U(2)$ group (the first two equations with $m_3=0$,
$m_1\neq m_2$), and Eqs.~(15), the mass parameters $M^2$, $\mu$
are given by
\[
M^2=\frac{2f^2 I_1 \left(m_1^3 +m_2^3 \right)}{m_1^2
\left(2m-m_1-m_2 \right)} ,\hspace{0.3in}\mu^2=-\frac{2g^2 I_1
\left(m_1^2 +m_1 m_2 +m_2^2 \right)}{m_1^2}.
\]
At the particular case $m=0$, one arrives from Eq.~(29) to the
relation $m_{00} =m_1+m_2$. If masses of fermions equal,
$m_1=m_2$, the mass matrix $m_{AB}$ is diagonal, and the result is
the mass of scalar bosons $m_{00}=2m_1$ and
$m_{11}=m_{22}=m_{33}=0 $. The mass $m_{00}$ is identified in the
quark models with the $\sigma$ meson mass \cite{Klevansky},
\cite{Volkov}, and fields $\Phi_a$ become the Goldstone massless
bosons because the symmetry $SU(2)$ is not broken. The same
relation holds in the four-fermion models without internal group
of symmetry \cite{Eguchi}, \cite{KruglovActa}, \cite{Nambu}.

To calculate the mass matrix $m_{AB}$ for the $U(3)$ group, it is
necessary to specify the solutions of the mass equation (22). If
we apply Lagrangian (1) to the real quarks interactions, one can
identify the triplet of fermions $\psi=(\psi_1, \psi_2,\psi_3 )$
with the triplet of light quarks $(u, d, s)$. In this case we may
take into consideration the real constituent quark masses (the
bare masses of quarks are the same, $m$) \cite{Volkov}
$m_1=m_u\approx m_2=m_d\approx 230$~MeV, $m_3 =m_s\approx
460$~MeV. Then one comes to the relation $m_3\approx m_1 +m_2$. It
should be stressed that the sign of the mass in the Dirac equation
for a fermion can be changed $m_0\rightarrow -m_0$ without loss of
generality. Therefore, we arrive at our case when $m_3=-m_1-m_2$
($m=f\Phi_0$). Then using the self-consistent equations (18) and
Eqs.~(15), one finds
\[
M^2=\frac{3f^2 I_1 \left(m_1^3 +m_2^3 +m_3^3 \right)}{m_1^2
\left(3m-m_1-m_2 -m_3 \right)} ,\hspace{0.3in}\mu^2=-\frac{2g^2
I_1 \left(m_1^2 +m_1 m_2 +m_2^2 \right)}{m_1^2}.
\]
With the aid of these equations, we find from Eqs.~(25) the
elements of the mass matrix:
\[
m_{00}^2=2\left(m_1^2+m_2^2+m_3^2\right)+\frac{2\left(m_1^3+m_2^3+m_3^3\right)}{3m-m_1-m_2-m_3}
\]
\[
=4\left(m_1^2 +m_2^2 +m_1 m_2\right)-\frac{2m_1 m_2 \left(m_1
+m_2\right)}{m} ,
\]
\[
m_{11}^2=m_{22}^2= m_{44}^2=m_{55}^2 =
 m_{66}^2=m_{77}^2 =0  ,
\]
\vspace{-7mm}
\begin{equation}\label{30}
\end{equation}
\vspace{-7mm}
\[
m_{88}^2=4m_3^2-\left(m_1+m_2\right)^2=3\left(m_1 +m_2\right)^2
,\hspace{0.3in}m_{03}^2=\sqrt{6}\left(m_1^2-m_2^2\right) ,
\]
\[
m_{08}^2=\sqrt{2}\left(m_1^2+m_2^2-2m_3^2\right)=-\sqrt{2}\left(m_1^2
+4m_1 m_2 +m_2^2\right) ,
\]
\[
m_{38}^2=\sqrt{3}\left(m_1^2-m_2^2\right) ,
\hspace{0.3in}m_{33}^2=\left(m_1-m_2\right)^2 .
\]
We imply here that $m\neq 0$. If the bare mass of fermions $m=0$,
then in accordance with the gap equations (18), $\Phi_0 \neq 0$,
and one arrives at the inequality (see Eqs.~(16)) $m_3\neq -m_1
-m_2$.

To obtain the mass spectrum of the bosonic fields $\Phi_A (x)$,
one must diagonalize the mass matrix $m_{AB}$. It follows from
Eqs.~(29), (30) that masses of the fields $\Phi_1 (x)$,  $\Phi_1
(x)$ vanish, which is in agreement with the Goldstone theorem
\cite{Goldstone} concerning spontaneous (or dynamical) symmetry
breaking. The other fields acquire nonzero masses. Consider the
case of the $U(2)$ group. To diagonalize the matrix with the
elements (29), we make the $SO(2)$-transformations
\begin{equation}
\Phi'_0 (x)=\Phi_0 (x)\cos\alpha -\Phi_3 (x)\sin\alpha
,\hspace{0.2in}\Phi'_3 (x)=\Phi_0 (x)\sin\alpha +\Phi_3
(x)\cos\alpha , \label{31}
\end{equation}
where
\[
\tan2\alpha=\frac{2m_{00}^2}{m_{33}^2-m_{00}^2}.
\]
Thus the mass matrix is diagonalized, and one comes to the
following masses of bosonic fields $\Phi'_0 (x)$, $\Phi'_3 (x)$
(the fields $\Phi_1 (x)$, $\Phi_2 (x)$ remain massless):
\[
m_{00}'^2=m_{00}^2\cos^2 \alpha+m_{33}^2 \sin^2 \alpha
-m_{03}^2\sin 2\alpha ,
\]
\vspace{-7mm}
\begin{equation}
\label{32}
\end{equation}
\vspace{-6mm}
\[
m_{33}'^2=m_{00}^2\sin^2 \alpha+m_{33}^2 \cos^2 \alpha +m_{03}^2
\sin 2\alpha .
\]
It should be noted that the transformations of the collective
fields (31) are generated by the corresponding fermion fields
$\psi (x)$. To diagonalize the mass matrix (30) for the group
$U(3)$, it is necessary to make the transformation of the fields
$\Phi_0 (x)$, $\Phi_3 (x)$, $\Phi_8 (x)$.

\section{Effective Lagrangian of Bosonic Fields}

Now we use the expression for the effective action
\cite{Goldstone}:
\[
S_{eff}=-\frac{1}{2}\int d^4 xd^4 y \Phi_A
(x)\Delta^{-1}_{AB}(x,y)\Phi_B (y)
\]
\begin{equation}
+\frac{1}{3!}\int d^4 xd^4 yd^4 z \Phi_A (x)\Phi_B(y)\Phi_C
(z)\Gamma_{ABC}(x,y,z)
 \label{33}
\end{equation}
\[
+\frac{1}{4!}\int d^4 xd^4 yd^4 zd^4 t \Phi_A (x)\Phi_B(y)\Phi_C
(z)\Phi_D(t)\Gamma_{ABCD}(x,y,z,t).
\]
Let us evaluate the terms in the sum (23), implying that
parameters of expansion $g^2/4\pi^2<1$, $f^2/4\pi^2<1$. We count
the components in (23) with $n=3$ and $n=4$. The fermion loops at
$n>4$ give small convergent expressions at
$\varepsilon\rightarrow0$ ($\Lambda\rightarrow \infty$). Vertex
functions entering Eq.~(33) are defined  as
\[
\Gamma_{ABC}(x,y,z)=\frac{\delta^3 S[\Phi]}{\delta\Phi_A
(x)\delta\Phi_B (y)\delta\Phi_C (z)},
\]
\vspace{-7mm}
\begin{equation}
\label{34}
\end{equation}
\vspace{-7mm}
\[
\Gamma_{ABCD}(x,y,z,t)=\frac{\delta^4 S[\Phi]}{\delta\Phi_A
(x)\delta\Phi_B (y)\delta\Phi_C (z)\delta\Phi_D (t)} .
\]
In the momentum space, three- and four-point functions are given
by
\[
\Gamma_{ABC}(k_1,k_2)=g_A g_B g_C i\mbox{tr}\biggl\{\int\frac{d^4
p}{(2\pi)^4}\biggl[ S_f (p+k_1-k_2)T_A S_f (p)T_B S_f(p+k_1)
\]
\vspace{-7mm}
\begin{equation}
\label{35}
\end{equation}
\vspace{-7mm}
\[
+S_f (p-k_1)T_B S_f (p)T_A S_f(p+k_2 -k_1)\biggr]T_C\biggr\} ,
\]
\[
\Gamma_{ABCD}(k_1,k_2,k_3)=g_A g_B g_C g_D
i\mbox{tr}\biggl\{\int\frac{d^4 p}{(2\pi)^4}\biggl\{ S_f (p)T_D
S_f (p+k_2)
\]
\[
\times \biggl[T_C S_f(p+k_2-k_3)T_B S_f (p-k_1)T_A +T_C S_f
(p+k_2-k_3)T_A S_f (p+k_1+k_2-k_3)T_B
\]
\vspace{-7mm}
\begin{equation}
\label{36}
\end{equation}
\vspace{-7mm}
\[
+ T_B S_f(p-k_1+k_3 )T_C S_f (p-k_1)T_A+T_A S_f (p+k_1+k_2)T_C S_f
(p+k_1+k_2-k_3)T_B
\]
\[
+T_A S_f(p+k_1 +k_2)T_B S_f(p +k_3)T_C+T_B S_f(p-k_1+k_3 )T_A S_f
(p+k_3)T_C \biggr]\biggr\}\biggr\},
\]
Inserting the Green function (14) (for $n=2$) into integrals (35),
(36), after calculations and renormalization with the help of
Eq.~(26), we arrive, for the case of $U(2)$ group ($T_a=\tau_a$),
at the effective Lagrangian (corresponding to the action (33)) of
interacting bosonic fields:
\[
{\cal L}_{int}(x) =-3g\left(m_1+m_2\right)\Phi_0 (x)\Phi_a^2
(x)-g\left(m_1+m_2\right)\Phi_0^3 (x)
\]
\vspace{-7mm}
\begin{equation}
\label{37}
\end{equation}
\vspace{-7mm}
\[
-3g\left(m_1-m_2\right)\Phi_3 (x)\Phi_0^2
(x)-g\left(m_1-m_2\right)\Phi_3 (x)\Phi_a^2 (x)-\frac{g^2}{4}
\mbox{tr}\Phi^4 (x) ,
\]
where $\Phi (x)=\Phi_0 (x)+\tau^a\Phi_a (x)$. If the vacuum field
$\Phi_3=0$, then according to Eq.~(29), the equality $m_1 =m_2$ is
valid and the symmetry of the group $U(2)$ is recovered.
Therefore, all fields $\Phi_a (x)$ become massless, but the field
$\Phi_0 (x)$ is still massive. In the same manner, one can
calculate the effective Lagrangian for the case of the $U(3)$
symmetry group of fermion fields.

\section{Discussion}

We have just considered the mass formation and DSB in the $U(n)$
four-fermion models (for $n=2,3$) with two coupling constants on
the basis of the functional integration method and the
bosonization procedure. In one loop approximation the propagators
of fermions  and collective Bose-fields (bound states of fermions)
have been evaluated. It is interesting that, in the framework of
the dimension regularization, a self-consistent consideration of
gap equations and condensates (vacuum averages of the collective
fields) leads to the mass formula (22) for the case of the $U(3)$
group. This new feature of four-fermion models allows us to
calculate the quark masses. However, it requires the consideration
of a model which possesses the additional colour symmetry, chiral
symmetry, and containing boson fields with quantum numbers of real
mesons (see \cite{Klevansky}, \cite{Volkov}). It is noted that
similar original relationships of the type (22) hold in the case
of the $U(n)$ ($n>3$) group (see \cite{Kruglov88} for the case of
$U(5)$ group). In our scheme the quadratic and logarithmic
diverging integrals, $I_j$ and $Z_3$ are connected by the relation
(27), but in the four-fermion models with the cutoff
regularization \cite{Volkov}, they are considered independent.
Using the dimension regularization, we found the masses of
collective boson fields $\Phi_A$ (29), (30) that are in agreement
with the Goldstone theorem. The original effective Lagrangian of
interacting scalar bosons (37) has also been derived for the case
$n=2$. So, a self-consistent consideration of four-fermion models
provides the mass relations for fermions and their bound states
(collective fields $\Phi_A (x)$).

It is emphasized that in the approach considered, the parameter of
the dimension regularization, $\varepsilon$, possesses physical
meaning because it enters the gap equations (18) which define mass
formulas. All integrals in such a scheme are finite leading to the
``finite renormalization". The four-fermion models can be
considered as an approximation to the description of the real
quark interactions. They lead to DSB but do not provide the
confinement of quarks. One may modify the model by introducing the
nonlocal interactions which approximate the linear potential
between quarks.


\begin{thebibliography}{99}

\bibitem{Weinberg} S. Weinberg, The Quantum Field Theory of
Fields, University Press, Vol.2, 1995.
\bibitem{Gasser} J. Gasser and H. Leutwyler, Nucl. Phys.
\textbf{B250} (1985), 465.
\bibitem{Buchvostov} A. P. Buchvostov, S. I. Kruglov, E. A. Kuraev. Chiral
Lagrangian and Physics of Mesons of Low-Energies. Published in
Material of the XXVIth LNPI Winter School, Elementary Particles
and Atom Nucleus Physics, Leningrad, pp. 52-123, 1991.
\bibitem{Klevansky}  S. P. Klevansky, Rev. Mod. Phys. {\bf 64} (1992), 649.
\bibitem{Volkov} M. K. Volkov, Elem. Chast. Atom. Yad. \textbf{17} (1986), 433;
Ann. Phys. \textbf{157} (1984), 282.
\bibitem{Eguchi} T. Eguchi, Phys. Rev. \textbf{D14} (1976), 2765;
C. Bender, F. Cooper, G. S. Guralnik, Ann. Phys. \textbf{109}
(1977), 165; K. Tamvakis and G. S. Guralnik, Phys. Rev.
\textbf{D18} (1978), 4551; A. Dhar, R. Shankar, S. R. Wadia, Phys.
Rev. \textbf{D31} (1985), 3256; D. Ebert and H. Reinhardt, Nucl.
Phys. \textbf{B271} (1986), 188.
\bibitem{KruglovActa} S. I. Kruglov, Acta Phys. Pol. \textbf{B15} (1984),
725.
\bibitem{Kruglov89} S. I. Kruglov, Sov. Phys. J. \textbf{32} (1989), 413
[Izv. VUZ. Fiz. No. 6 (1989), 5].
\bibitem{Shuryak}  E. V. Shuryak, Phys. Lett. \textbf{B107} (1981),
103; Nucl. Phys. \textbf{B203} (1982), 93; \textbf{B214} (1983),
237.
\bibitem{Dyakonov} D. I. Dyakonov, V.Yu. Petrov, Nucl. Phys. \textbf{B245}
(1984), 259; \textbf{B272} (1986), 457.
\bibitem{Hooft}  G. 't Hooft, Phys. Rev. Lett. \textbf{37} (1976), 8;
Phys. Rev. \textbf{D14} (1976), 3432.
\bibitem{Kruglov90} S. I. Kruglov, Acta Phys. Pol. \textbf{B21} (1990),
985; Sov. Phys. J. \textbf{33} (1990), 580 [Izv. VUZ. Fiz. No. 7
(1990), 36].
\bibitem{Nambu} Y. Nambu and G. Jona-Lasinio, Phys. Rev. \textbf{122} (1961), 345; \textbf{124}
(1961), 246; V. G. Vaks, A. I. Larkin, Zh. Eksp. Teor. Fiz.
\textbf{40} (1961), 282 [Sov. Phys. JETP \textbf{13} (1961), 192];
B. A. Arbusov, A. N. Tavchelidze, R. N. Faustov, Dokl. Akad. Nauk
SSSR, \textbf{139} (1961), 345 [Sov. Phys. Doklady \textbf{6}
(1962), 598];
\bibitem{Nambu60} Y. Nambu, Phys. Rev. Lett. \textbf{4} (1960),
380.
\bibitem{KruglovTMF} S. I. Kruglov, Theor. Math. Phys.
\textbf{84} (1990), 945 [Teor. Mat. Fiz. \textbf{84} (1990), 388].
\bibitem{Simonov}  Yu. A. Simonov, Usp. Fiz. Nauk \textbf{166} (1996), 337;
Phys. Rev. \textbf{D65} (2002), 094018.
\bibitem{Kruglov} S. I. Kruglov, Phys. Lett. B {\bf 390}, 283
(1997); Phys. Lett. B {\bf 397} (1997), 283; Phys. Rev. D {\bf 60}
(1999), 116009; arXiv:hep-ph/0110101.
\bibitem{Kruglov84} S. I. Kruglov, Vestsi Akad. Navuk BSSR, ser. fiz.-mat. navuk
[Proceedings of the National Academy of Sciences of Belarus,
Series of Physical- Mathematical Sciences] No. 6 (1985), 87 [in
Russian].
\bibitem{Kruglov85} S. I. Kruglov, Dokl. Acad. Nauk BSSR
[Doklady of the Academy of Sciences of Belarus] \textbf{29}
(1985), 42 [in Russian].
\bibitem{Pervushin} V. P. Pervushin, H. Reinhardt, D. Ebert,
Elem. Chast. At. Yad. \textbf{10} (1979), 1114.
\bibitem{Jona} G. Jona-Lasinio, Nuovo Cim. \textbf{34} (1964),
1790.
\bibitem{Coleman} S. Coleman and E. Weinberg, Phys. Rev. \textbf{D7} (1973), 1888.
\bibitem{Hooft72} G. 't Hooft and M. Veltman, Nucl. Phys. \textbf{B44}
(1972), 189.
\bibitem{Ramond} P. Ramond, Field Theory, Benjamin-Cummings,
Reading MA (1981).
\bibitem{Scadron} R. Delbourgo and M. D. Scadron, Mod. Phys. Lett.
\textbf{A10} (1995), 25.
\bibitem{Goldstone} J. Goldstone, A. Salam, and S. Weinberg, Phys. Rev. \textbf{127} (1962),
965.
\bibitem{Kruglov88} S. I. Kruglov, Sov. Phys. J. \textbf{31} (1988),
198 [Izv. VUZ. Fiz. No. 3 (1988), 30].

\end{thebibliography}
\end{document}